\documentclass[preprint2,twoside]{hwo}

\usepackage{lipsum}

\def\stitle{HWO Astrometry for Exoplanets and Dark Matter}                    
\def\sauthors{F.~Malbet, J.~Amiaux, F.~Ardellier-Desages et al.}

\input{hwo.h}

\begin{document}

\title{\textbf{\LARGE Very High Precision Astrometry for Exoplanets and Dark Matter with the Habitable Worlds Observatory}}
\author{\textbf{\large Fabien Malbet,$^1$ Jérôme Amiaux,$^2$ Florence
    Ardellier-Desages,$^2$ Éric Doumayrou,$^2$ Pierre-Antoine
    Frugier,$^2$ Renaud Goullioud,$^3$ Thomas Greene,$^4$ Lucas
    Labadie,$^5$ Pierre-Olivier Lagage,$^2$ Manon Lizzana,$^{1,6,7}$ Alain
    Léger,$^8$ Thierry Lépine,$^9$ Gary Mamon,$^{10}$ Jérôme
    Martignac,$^2$ Fabrice Pancher,$^1$ Thibault Pichon,$^2$ Aki
    Roberge,$^{11}$ Samuel Ronayette,$^2$ Hugo Rousset,$^1$ Sébastien
    Soler,$^1$ Alessandro Sozzetti,$^{12}$ Thierry Tourette$^2$}}
\affil{$^1$\small\it Univ.\ Grenoble Alpes, CNRS, IPAG, Grenoble, France}
\affil{$^2$\small\it Univ. Paris-Saclay, CEA, Saclay, France}
\affil{$^3$\small\it Jet Propulsion Laboratory, California Institute of  Technology, Pasadena, CA, USA}
\affil{$^4$\small\it IPAC, California Institute of  Technology, Pasadena, CA, USA}
\affil{$^5$\small\it Univ.\ of  Cologne, Cologne, Germany}
\affil{$^6$\small\it Univ.\ Paris-Saclay, CNRS, Institut d'astrophysique spatiale, Orsay, France}
\affil{$^7$\small\it Centre National d'Études Spatiales, Toulouse, France}
\affil{$^8$\small\it Pyxalis, Moirans, France}
\affil{$^9$\small\it Institut d'Optique \& Hubert Curien Lab, Univ.\ de Lyon, Saint-Etienne, France}
\affil{$^{10}$\small\it Sorbonne Université, CNRS, Institut d’Astrophysique de Paris, Paris, France}
\affil{$^{11}$\small\it NASA Goddard Space Flight Center, Greenbelt, MD, USA}
\affil{$^{12}$\small\it Obs.\ Torino/INAF, Pino Torinese, Italy}



\begin{abstract}
  Astrometry, one of the oldest branches of astronomy, has been
  revolutionized by missions like Hipparcos and especially Gaia, which mapped
  billions of stars with extraordinary precision. However, challenges
  such as detecting Earth-like exoplanets in nearby habitable zones
  and probing the influence of dark matter in galactic environments
  require sub-microarcsecond accuracy.

  With a 6–8 meter large-aperture telescope operating across at
  visible wavelengths, the Habitable Worlds Observatory by NASA can
  combine astrometry and direct imaging to detect rocky exoplanets
  within 10 parsecs and study their atmospheres. We consider here the
  scientific requirements and present a concept for a dedicated
  astrometric instrument on HWO. It is capable to produce
  diffraction-limited images of large fields, achieving a point-spread
  function (PSF) precision of 20 milliarcseconds. Equipped with a
  detector calibration system, HWO can perform high precision
  astrometry, and, detect and measure the orbit of Earth-mass planets
  in the habitable zone of Nearby Solar-type stars. HWO can
  dramatically improve current constraints on the self-interaction
  cross section of heavy dark matter particles (WIMPs) and on the
  masses of ultra-high dark matter particles, through the study
 of stellar motions in galactic environments.

  The visible channel of the instrument features a large CMOS-based
  focal plane with stitched pixel arrays, enabling a large field of
  view. The ``Detector Calibration Unit'' system uses interferometric
  laser fringes to calibrate pixel positions. Using differential
  astrometry and pointed observations with a stable telescope design
  enables extended integration times, enhancing sensitivity to
  sub-microarcsecond precision for detecting exoplanets and studying
  dark matter through stellar motion.

\bigskip
\end{abstract}

\section{Introduction}

\begin{figure*}[t]
  \centering
  \includegraphics[width=0.95\textwidth]{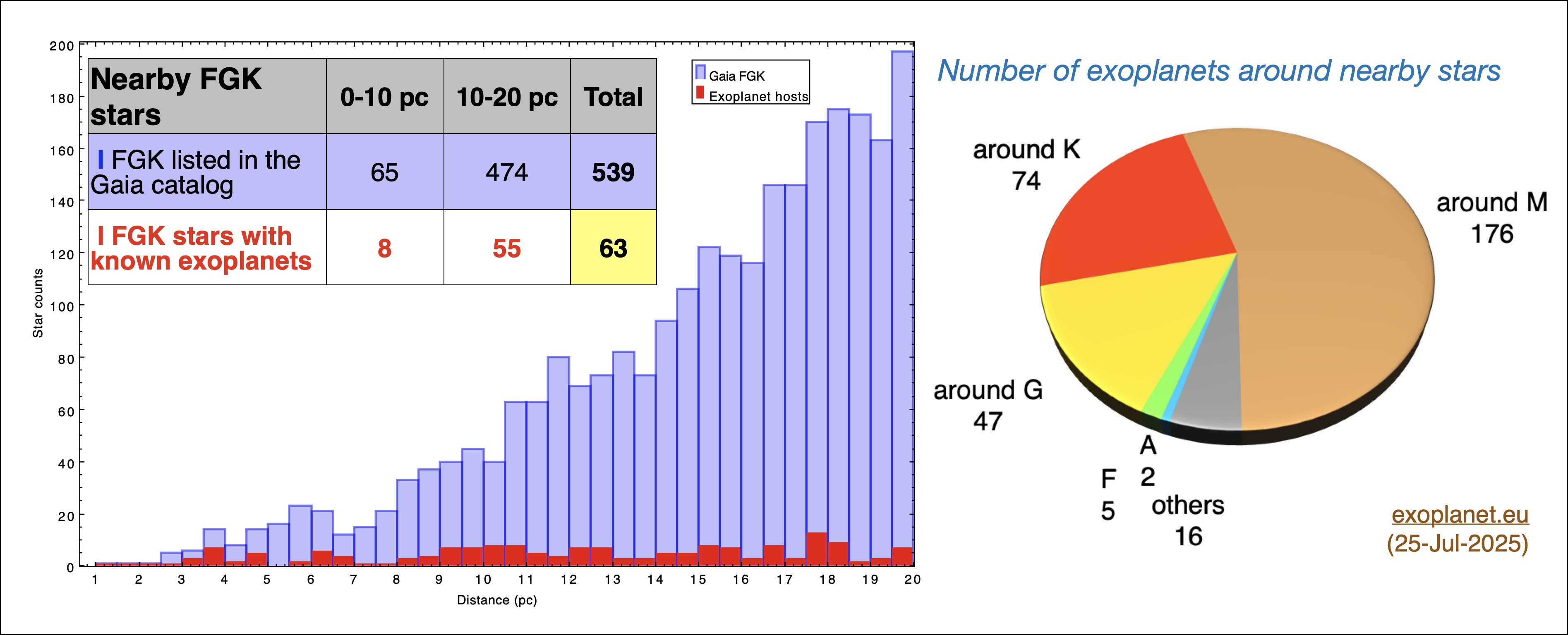}
  \caption{Exoplanets known around nearby solar-type stars. Left panel
    : number of nearby stars ($D\leq 20\,\mbox{pc})$ with (in red) or
    without (in blue) known exoplanets exxtracted from the exoplanets
    encyclopedia at the date of 25 July 2025. Only 12\% of nearby
    stars have known exoplanets so far and mostly gaseous ones! Right
    panel : number of exoplanets found around nearby stars of
    different spectral type around nearby stars (data from The
    Extrasolar Planets Encyclopaedia).}
  \label{fig:nearby-exoplanets}
\end{figure*}
\begin{figure*}[t]
  \centering
  \includegraphics[width=0.95\textwidth]{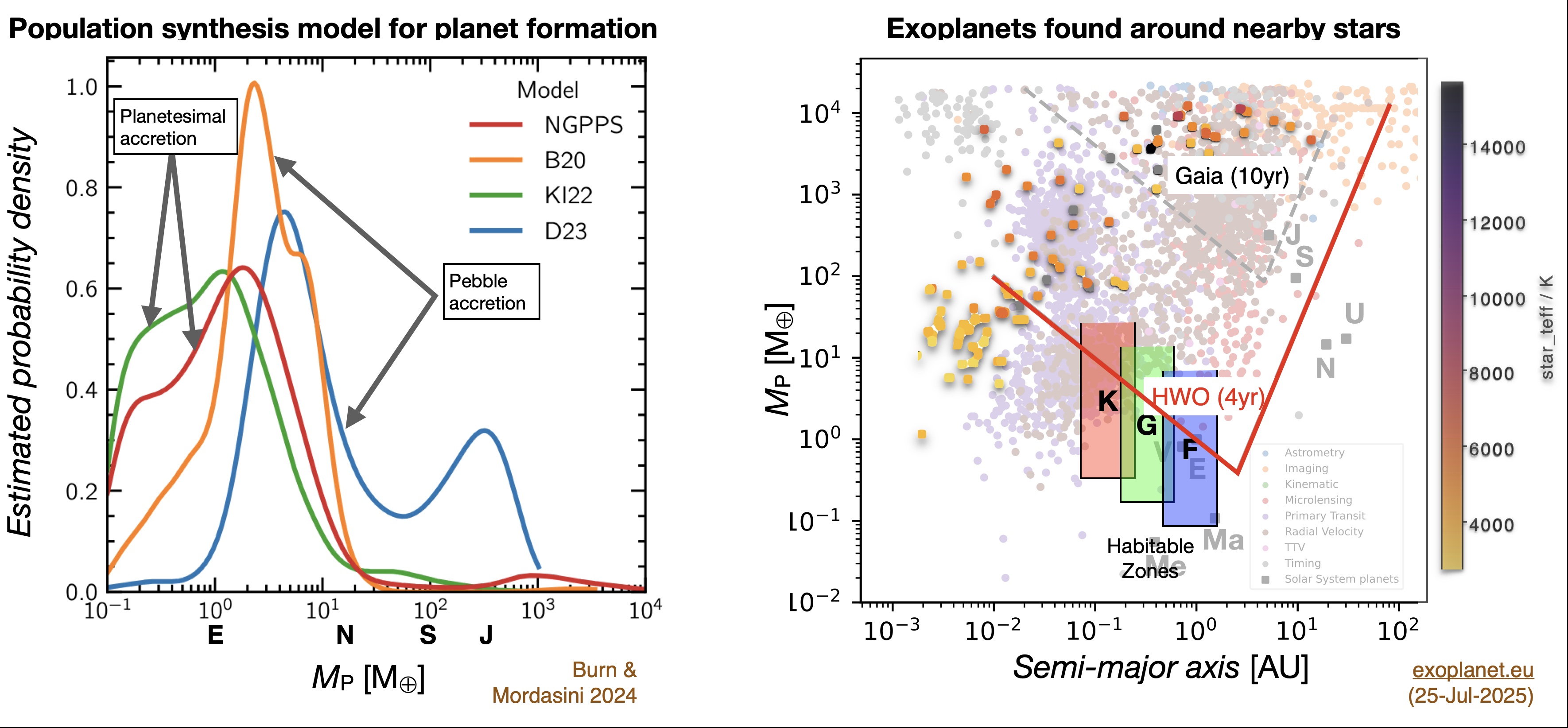}
  \caption{The low-mass planetary components that is currently
    missing. Left panel : Kernel density estimates of four synthetic
    planetary populations \citep[from][]{Burn&Mordasini24}. Right
    panel: mass versus semi-major axis parameter space for known
    exoplanets in function of the detection technique. The nearby
    exoplanets ($D\leq 20\,\mbox{pc}$ are displayed in clear above the
    full sample of known exoplanets at the date of 25 July 2025 (data
    from The Extrasolar Planets Encyclopaedia).}
  \label{fig:low-mass-exoplanets}
\end{figure*}

Astrometry is one of the most fundamental techniques in astronomy,
tracing its heritage back to the earliest stellar catalogs. Over the
past decades, space missions such as Hipparcos and Gaia have delivered
a revolution in our knowledge of stellar positions, parallaxes, and
proper motions. Gaia in particular has measured over two billion
stars, opening new windows on stellar populations, galactic dynamics,
and planetary systems. Yet, even Gaia's exquisite precision (down to
tens of microarcseconds) is insufficient for detecting Earth-mass
planets in the habitable zones of nearby stars or for probing subtle
dark matter signatures in dwarf galaxies and galactic halos.

The Habitable Worlds Observatory (hereafter HWO), despite initially not being
designed to address these challenges, has the potential to achieve
measurements with accuracies approaching $0.3\,\mu\mathrm{as}$ by incorporating a
dedicated high-precision astrometric instrument. This instrument will
enable the detection of true Earth analogs at 10\,pc and the
dynamical mapping of dark matter distributions on both small and large
scales. This article develops the scientific rationale, instrumental
principles, anticipated performance, and synergies with alternative
methodologies.

\section{Science objectives}

The U.S. decadal survey on astronomy and astrophysics for the 2020s \emph{Pathways
  to Discovery in Astronomy and Astrophysics for the 2020s}
\citep{Decadal2020} identified the detection and characterization of
Earth-like exoplanets and the study of dark matter as top priorities
for the coming decades. Similarly, the European Space Agency's
long-term science plan \emph{Voyage 2050} \citep{Voyage2050}
emphasized the search for habitable worlds and the understanding of
the nature of dark matter as central science themes.

In particular, Section 3.1.7 of the ESA Voyage 2050 report stresses
that the next steps in space astrometry should either aim at improving
relative astrometric accuracy by an order of magnitude or extend
global astrometry to the near-infrared domain. The report highlights
that high precision relative astrometry at the sub-microarcsecond
level (down to 0.2 $\mu$as) would enable two transformative
breakthroughs: probing the distribution of dark matter in galactic
environments, and detecting and fully characterizing the orbital
architectures of planetary systems hosting habitable planets around
the nearest stars. Achieving such capabilities requires however
substantial advances in detector technology, metrology, and spacecraft
stability.

\subsection{Exoplanet census of nearby FGK stars}

A central motivation for HWO astrometry is the discovery and mass
determination of rocky planets in the habitable zones of the nearest
Sun-like stars. Only about 12\% of nearby FGK stars currently have
known exoplanets, mostly giant planets
(Fig.~\ref{fig:nearby-exoplanets}). The low-mass, temperate planets
remain elusive, yet they are the most compelling targets for
biosignature searches. Astrometry provides complete orbital solutions
and absolute masses, overcoming limitations of radial velocity (RV)
techniques \citep{Meunier+20, Meunier&Lagrange22}. It is especially powerful for planets on longer orbits,
inaccessible to transit methods.

To meet the Decadal Survey goal of obtaining $\sim25$ atmospheric
spectra of potentially habitable exoplanets, the HWO strategy combines
a large space telescope equipped with a high-contrast coronagraph, an
optimized target list of about 100–160 nearby FGK stars, and yield
models based on occurrence rates ($\sim0.24$ Earth-like planets per
star in the habitable zone). By directly imaging planets around these
stars and allocating sufficient integration time with repeated visits,
the mission can reach the signal-to-noise needed for spectral
characterization. Achieving this requires maturing technologies for
starlight suppression, wavefront stability, and ultra-stable
detectors, while precursor observations refine target selection and
reduce risk. Together, these elements are designed to ensure a robust
sample of $\sim25$ spectra of rocky, temperate exoplanets suitable for
habitability studies. Such a target list would significantly reduce
the time spent observing for the spectroscopic study.

An astrometric survey targeting 100--200 of the nearest FGK stars can
yield dozens of Earth-mass planets. These detections will not only
establish occurrence rates but can also deliver the prioritized target
list for HWO's coronagraphic and spectroscopic follow-up. Stellar
jitter, often a limiting factor for RV, is less problematic for
astrometry at these scales, typically contributing $<0.1\,\mu$as for
Sun-like stars at 10 pc. This places the exoplanet detection threshold
securely within reach for Earth analogues.

The combination of astrometry, RV, and direct imaging is
transformative. RVs supply high cadence but ambiguous $M\sin i$
estimates, resolved by astrometry. Direct imaging reveals atmospheric
spectra but requires orbital ephemerides for efficient scheduling,
provided by astrometry. Together, the three approaches enable the most
complete and accurate characterization of exoplanetary systems.

\subsection{Probing dark matter with proper motions}

Dark matter (DM) dominates the matter content of the Universe, but its
nature remains unknown, in particular the mass of the DM particle and
the possible cross-section for DM annihilation through self
interaction. DM halos play a central role in shaping the
dynamics of galaxies and clusters. Their internal density profiles and
subhalo populations remain key open questions in cosmology. Dwarf
spheroidal galaxies, being strongly dark-matter dominated, provide
fundamental laboratories to test whether halos are cored or cuspy,
thus directly probing predictions of the cold dark matter (CDM)
paradigm and its alternatives, such as self-interacting or fuzzy dark
matter. CDM simulations also predict the existence of numerous dark
subhalos devoid of stars, which may nevertheless be revealed through
their gravitational perturbations on stellar streams and the
kinematics of stars in the Milky Way disk.

High-precision astrometry with HWO, reaching sub-km/s proper-motion
accuracy at the distances of dwarf galaxies, offers the means to
distinguish cusp from DM core profiles, detect the dynamical
signatures of otherwise invisible subhalos, and constrain the overall
shape of the Milky Way’s halo using tracers such as hypervelocity
stars. Together, these three avenues—inner density slopes, subhalo
detection, and halo shape—illustrate the transformative potential of
microarcsecond astrometry with HWO for advancing our understanding of
dark matter. HWO would provide a major improvement on the constraints
on the fuzzy DM particle mass if DM is ultra-light, as well as on the
cross section of self-interaction if DM is heavy.

\section{Measurement principle and instrument concept}
\begin{figure*}[t]
  \centering
  \includegraphics[width=0.95\textwidth]{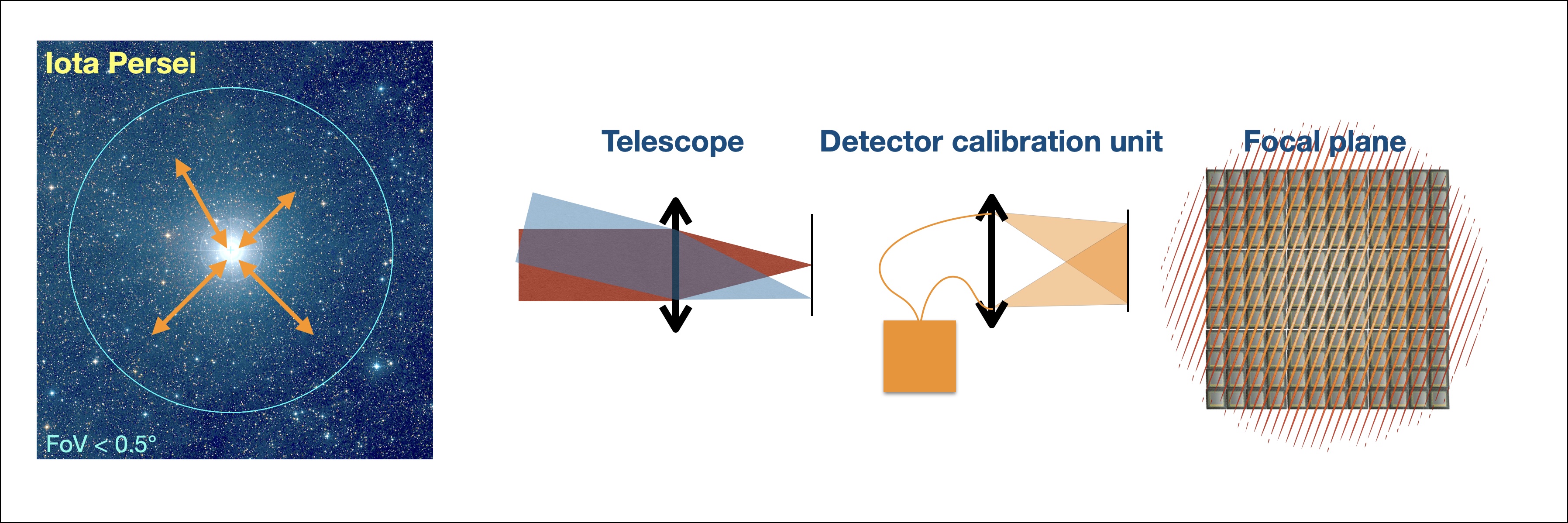}
  \caption{Principle of relative differential astrometry. Left panel: this panel presents an image of the
    0.5-degree field of view of the exoplanet stellar host  \emph{iota
      Persei} (from DSS). The arrows show the distances between the
    target central bright star and some reference stars.
    Right panel: instrument principle which includes a telescope which
    is diffraction-limited allowing apparent angle measurements, a
    detector calibration unit that projects interferometric fringes
    created by the interferences between laser light onto the focal
    detector.}
  \label{fig:principle}
\end{figure*}
\begin{figure*}[t]
  \centering
  \includegraphics[width=0.95\textwidth]{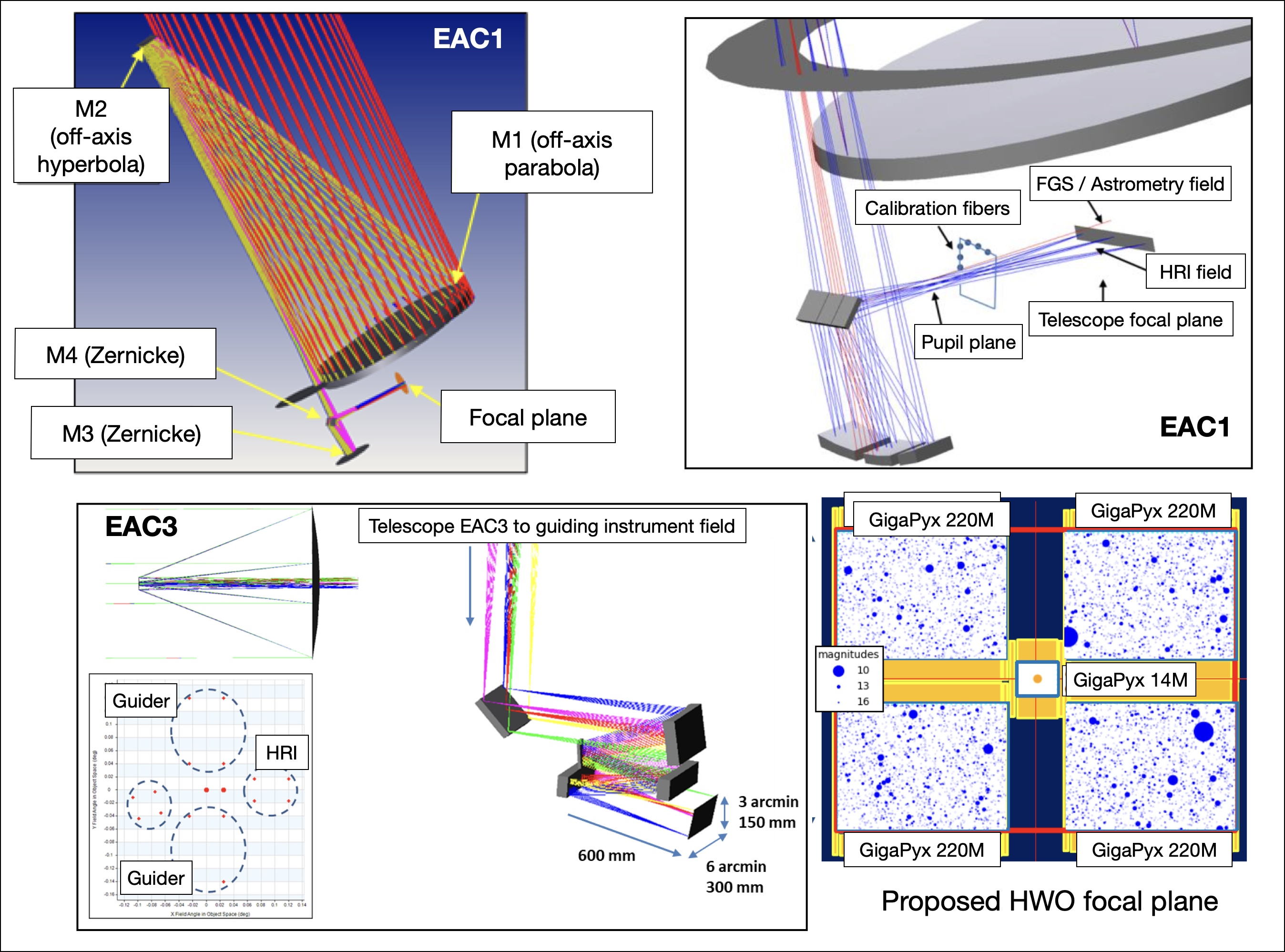}
  \caption{A possible Focal Plane dedicated instrument for HWO. Upper
    left panel: HWO  telescope configuration for Exploratory
    Architecture Concepts) EAC1. Upper right panel : location of the
    focal plane and the detectur Calibration unit in the EAC1 case.
    Bottom left panel : location of the focal plane and the detector
    Calibration unit in the EAC3 case. Bottom right panel : possible
    focal plane concept with 4 GigaPyx 220M detectors and a central
    GigaPyx 14M detector.}
  \label{fig:implementation}
\end{figure*}

The astrometric capability envisioned for the Habitable Worlds
Observatory relies on the principle of high-precision differential
astrometry (left part of Fig.~\ref{fig:principle}). In practice, the science target
is placed at the center of the field of view, and its angular position
is measured relative to an ensemble of reference stars distributed
across the detector. By repeating these measurements at multiple
epochs over the mission lifetime, it is possible to disentangle the
different contributions to the observed motion (parallax, proper
motion, and orbital reflex motion induced by exoplanets) with
microarcsecond precision. The differential approach minimizes
common-mode errors introduced by the telescope optics or spacecraft
instabilities and is therefore particularly well-suited to HWO space-based
observations.

To enable such accuracy, the instrument concept (right part of
Fig.~\ref{fig:principle}) centers on a large visible-light CMOS-based
focal plane composed of stitched pixel arrays
(Fig.~\ref{fig:implementation}), providing both high stability and a
wide field of view. The point-spread function (PSF) is expected to be
sampled at a the diffraction-limited resolution of roughly 20
milliarcseconds, ensuring Nyquist sampling in the blue optical regime
where photon fluxes from Sun-like stars are maximized. Achieving
long-term stability requires a thermal and mechanical design of the
telescope optimized to suppress slow drifts and minimize variations in
the optical train that could bias the astrometric solution.

A important element of the concept is the Detector Calibration Unit
(DCU). This subsystem projects interferometric laser fringes across
the focal plane, enabling sub-pixel calibration of detector geometry
and intra-pixel response. By continuously monitoring pixel positions
at the nanometer scale, the DCU ensures that systematic uncertainties
associated with the detector remain below the mission noise floor.
Combined with stable spacecraft pointing and advanced wavefront
control, this calibration strategy allows the focal plane to operate
at the level of a few tenths of a microarcsecond over years of
integration.

Finally, the observing strategy itself contributes to the robustness
of the measurements. By alternating the orientation of the telescope
on the sky, sampling different reference star ensembles, and
revisiting each target multiple times, systematic errors can be
averaged down. The combination of instrumental design, calibration
techniques, and optimized observing cadence is what makes it possible
to achieve the sub-microarcsecond performance required to detect
Earth-mass planets at 10 parsecs and to measure the subtle proper
motions associated with dark matter studies.

Further information regarding the astrometry instrument that has been
proposed for HWO can be found in this volume : \citet{Amiaux+25}
present a system analysis for an instrument on-board HWO dedicated to
high-precision high-accuracy differential astrometry with the main
science driver to detect Earth-like planets signal around the closest
Sun-like stars; and \citet{Lizzana+25} present a characterization of
the 46-megapixel Gigapyx detector from Pyxalis, which is an
appropriate detector for an astrometric instrument, explain how to use
a testbed to measure pixel positions, and, present a method for
adjusting the telescope optical distortion which was tested using an
optical bench developed at IPAG in France.

\section{Expected performance and noise budget}
\begin{table*}[t]
  \centering
  \includegraphics[width=0.95\textwidth]{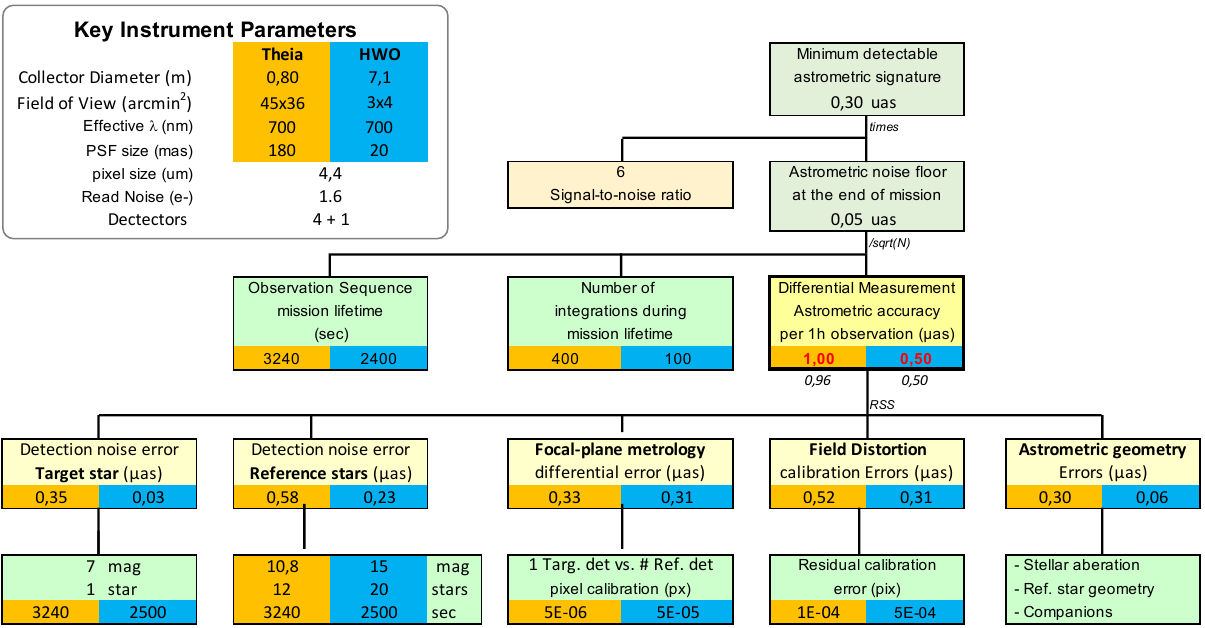}
  \caption{Astrometric error budget for bright targets with Theia
    \cite[0.8-m telescope with 0.5\,deg FoV, in orange color, see ][]{Malbet+22} and 
    HWO (7-m telescope with $3\times2$ arcmin$^2$ FoV, in blue color). }
  \label{tab:error-budget}
\end{table*}

To detect an Earth-mass planet at 10 pc requires sensitivity at the
$0.3\,\mu$as level. For other planets the astrometric signal
scales as :
\begin{equation}
  \label{eq:1}
  \alpha = 0.33  \,
  \left(\frac{M_\mathrm{P}}{1\,M_\oplus}\right)
  \left(\frac{a_\mathrm{P}}{1\,\mathrm{AU}}\right) 
  \left(\frac{M_\star}{1\,M_\odot}\right)^{-1}
  \left(\frac{D}{10\,\mathrm{pc}}\right)^{-1} 
  \hspace{-1ex}
  \mu\mathrm{as}
\nonumber
\end{equation}
where $M_\mathrm{P}$ is the planet mass, $a_\mathrm{P}$ is the planet
semi-major axis, $M_\star$ is the star mass and $D$ is the distance
from the observer.

Achieving this unprecedented precision of 0.3\,$\mu$as
required for the Habitable Worlds Observatory (HWO) demands a rigorous
understanding and control of all potential sources of error. The
astrometric error budget for HWO is a complex interplay of
instrumental, astrophysical, and environmental factors, each
contributing to the overall uncertainty in the measurements. At the
heart of the instrumental errors are optical aberrations, which arise
from imperfections in the telescope’s optics and misalignments in the
optical path.

To achieve the detection of Earth-mass exoplanets with a
signal-to-noise ratio (SNR) of 6, the Habitable Worlds Observatory
(HWO) must attain the residual error of 0.3\,$\mu$as after averaging
controlled at the level of 0.05\,$\mu$as. This stringent requirement
is met by accumulating approximately 100 individual measurements, each
with an uncertainty of 0.5\,$\mu$as, over the mission’s 3- to 4-year
duration.

However, this approach assumes that random errors dominate and these
errors are uncorrelated between measurements. Systematic errors—such
as long-term instrumental drifts, calibration biases, or residual
optical distortions—must be independently controlled to avoid
degrading the final precision. HWO addresses these challenges through
interferometric calibrations and repeated observations of stable
reference star fields to minimize systematic biases.

Table~\ref{tab:error-budget} summarizes the expected contributions of
each error source for HWO and for Theia, which is a 0.8-m telescope
dedicated to astrometry that is proposed to ESA \citep{Malbet+22}.
Theia is a project for a dedicated astrometry telescope. Its
advantages include more exposures and exposure time, and, a much wider
field of view. However, it also has disadvantages, such as lower
sensitivity and poorer spatial resolution. The table demonstrates how
the mission’s design accounts for each potential uncertainty. This
comprehensive approach to error management will be validated through a
combination of laboratory tests, simulations, and in-flight
calibration. Pre-launch tests will characterize instrumental errors,
while simulations will model astrophysical and environmental
contributions. In-flight calibration, using known star fields
(globular clusters) and interferometric techniques, will verify the
error budget and ensure that HWO meets its precision goals.

Astrophysical errors, such as those caused by stellar activity,
binarity, and reference star instability, add another layer of
complexity. Stellar activity, including starspots and flares, can
cause apparent shifts in a star’s photometric center, leading to
astrometric errors.

The stability of reference stars is another critical factor in
differential astrometry. Any proper motion, parallax, or variability
in the reference stars can introduce errors in the measured positions
of the science targets. To minimize this, HWO will carefully select
reference stars based on their stability and known proper motions, and
will use a large ensemble of reference stars to average out individual
variations.

Detector noise is a significant source of instrumental error. The
Pyxalis CMOS GIGA220M detectors, while designed for low readout noise
and high quantum efficiency, still introduce uncertainties due to dark
current, pixel gains and readout noise. Calibration inaccuracies also pose a
challenge, as errors in the pixel scale or geometric distortion
corrections can propagate into the final astrometric measurements.
These inaccuracies can introduce systematic biases if not properly
managed. HWO’s approach to this issue involves a combination of
laboratory calibration, in-flight calibration using known star fields,
and interferometric techniques. These methods ensure that the
calibration remains accurate and stable throughout the mission,
reducing the risk of systematic errors.

Environmental errors, such as thermal fluctuations and mechanical
vibrations, further complicate the error budget. Thermal fluctuations
can cause expansions and contractions in the telescope structure,
leading to misalignments and changes in the optical path. These
effects can introduce errors if not controlled. HWO
will mitigate thermal fluctuations using active thermal control
systems and low thermal expansion materials, ensuring that the
telescope maintains a stable operating temperature. Mechanical
vibrations, caused by spacecraft operations or reaction wheel
activity, can introduce jitter in the telescope’s pointing.

In conclusion, the astrometric error budget for HWO is a critical
component of the mission’s design and operation. By meticulously
characterizing and mitigating each source of error, HWO astrometric
instrument can achieve the precision necessary to detect Earth-like
exoplanets and investigate the nature of dark matter.

\section{Conclusions and perspectives}

Astrometry with HWO represents a natural continuation of the Hipparcos
and Gaia legacy, pushing precision into the sub-microarcsecond domain.
It uniquely enables the detection of true Earth analogues around
nearby stars and delivers essential constraints on dark matter
structure in galaxies. The required technology is challenging but
achievable, and the scientific rewards, from identifying habitable
worlds to probing the fundamental properties of dark matter, are
profound. A dedicated astrometric instrument aboard HWO should
therefore be considered a strategic priority.

\section*{Acknowledgements}

We acknowledge the HWO community and the many contributors to concept
studies that motivated this work. FM thanks collaborators and
institutions listed in the author affiliations for support and
discussions.

With regard to the funding of our research, we also acknowledge the
support of the LabEx FOCUS ANR-11-LABX-0013 and the CNES agency. ML
would like to acknowledge the support of her PhD grant from CNES and
Pyxalis.
  
This research has made use of NASA Astrophysics Data System
Bibliographic Services and of data obtained from or tools provided by the portal \href{exoplanet.eu}{exoplanet.eu} of The Extrasolar
Planets Encyclopaedia.

\bibliography{HWO-astrometry-Malbet.bib}

\end{document}